\documentstyle[12pt]{article}
\begin{document}
\date{\today}
\pagestyle{plain}
\title{Compton effect in dielectric medium}
\author{Miroslav Pardy\\
Institute of Plasma Physics ASCR\\
Prague Asterix Laser System, PALS\\
Za Slovankou 3, 182 21 Prague 8, Czech Republic\\[3mm]
and\\[3mm]
Department of Physical Electronics \\
Masaryk University \\
Kotl\'{a}\v{r}sk\'{a} 2, 611 37 Brno, Czech Republic\\
e-mail:pamir@physics.muni.cz}
\date{\today}
\maketitle
\vspace{50mm}

\begin{abstract}
We determine the Compton effect from the Volkov solution of the Dirac
equation for a process in medium with the index of refraction $n$.
Volkov solution involves the mass shift, or, the mass renormalization
of an electron.
We determine the modified  Compton formula for the considered physical
situation. The index of refraction causes that the wave lengths of the
scattered photons are shorter for some angles than the wave lengths of
the original photons. This is anomalous Compton effect.
Since the wave length shift for the visible light is only
0,01 percent, it  means that the Compton
effect for the visible light in the dielectric medium
needs experimental virtuosity.

\end{abstract}

\vspace{2mm}


\vspace{2mm}


\newpage

\baselineskip 15 pt

\section{Introduction}

The process of the photon scattering on electron is called the Compton
process after Arthur Compton who made the first measurement of 
photon-electron scattering in 1922. He used the monochromatic X-rays
for the determination of their scattering on the free electrons in a block
of graphite. He observed that the scattered beam consisted of X-rays of a
longer wavelength in addition to the original beam. This was explained
successfully by applying the laws of conservation of energy and momentum in
the collision between X-ray photon and free electron inside graphite block.

He derived the equation

$$\lambda' - \lambda = \frac {h}{mc}(1 - \cos\theta),\eqno(1)$$
where $\lambda'$ is wavelength of the scattered X-ray and $\theta$ is the
angle between the incident and scattered X-ray.
The scattering was considered in the laboratory frame where electron was at
rest. The considered process was so called one photon process with the
symbolic equation

$$\gamma + e \rightarrow \gamma + e.\eqno(2)$$

At present time with the high power lasers there is possible so called the
multiphoton scattering according to equation:

$$N\gamma + e \rightarrow \gamma + e,\eqno(3)$$
or,

$$N\gamma + e \rightarrow M\gamma + e,\eqno(4)$$
where N and M are numbers of photons participating in the scattering.

The equation (3) is the symbolic expression of the two different physical
processes. One process is the nonlinear Compton effect in which several
photons are absorbed at a single point, but only single high-energy photon
is emitted. The second process is the interaction where electron scatters twice
or more as it traverses the laser focus. Similar interpretation is
 valid for equation (4).

In our article, the attention is
concentrated to the situation where the Compton process is considered in
dielectric medium with index of refraction $n$. The index of refraction is
generated microscopically by the system of bound electrons and this
system forms a medium called dielectric which enables transmitting of
electromagnetic waves in this medium. There are also  free electrons in this
medium. The scattering of photons of
the medium on free electrons in this medium then can occur.
At present time,
the beam of photons can be realized by lasers which
play prestige role in all physical laboratories.

Our aim is to determine the Compton process
in the dielectric medium as a result of the Volkov
solution of the Dirac equation. The Volkov solution involves not only the
one-photon scattering but also the multiphoton scattering of photons on
electron. In time of Compton experiment in 1922,
the Volkov solution was  not known, because the Dirac equation was
published in 1928 and the  Volkov solution in 1935 [1].

To be pedagogically clear, we remember in the next section some known
ideas concerning the Volkov solution of the Dirac equation.

\section{Volkov solution of the Dirac equation in vacuum}

Let us remember the derivation of the Volkov
solution of the Dirac
equation in vacuum (we use here the method of derivation and
metric convention of Berestetzkii {\it et al.}[2]):

$$(\gamma(p-eA) - m)\psi = 0, \eqno(5)$$
where

$$A^{\mu} = A^{\mu}(\varphi); \quad \varphi = kx.\eqno(6)$$

We suppose that the four-potential satisfies the Lorentz gauge condition

$$\partial_{\mu}A^{\mu} = k_{\mu}\left(A^{\mu}\right)' =
\left(k_{\mu}A^{\mu}\right)' = 0, \eqno(7)$$
where the prime denotes derivation with regard to $\varphi$. From the
last equation follows

$$kA = const = 0,\eqno(8)$$
because we can put the constant to zero. The tensor of electromagnetic field is

$$F_{\mu\nu} = k_{\mu}A'_{\nu} - k_{\nu}A'_{\mu}.\eqno(9)$$

Instead of the linear Dirac equation (5), we consider the quadratical
equation, which we get by multiplication of the linear equation by
operator $(\gamma(p-eA) + m)$, [2].
We get:

$$\left[(p - eA)^{2} -m^{2} - \frac{i}{2}eF_{\mu\nu}
\sigma^{\mu\nu}\right]\psi = 0. \eqno(10)$$

Using $\partial_{\mu}(A^{\mu}\psi) = A^{\mu}\partial_{\mu}\psi$, which
follows from eq. (7), and $\partial_{\mu}\partial^{\mu} =
\partial^{2} = -p^{2}
$, with $p_{\mu} = i(\partial /\partial x^{\mu}) = i\partial_{\mu}$, we get
the quadratical Dirac equation for the four potential of the plane wave:

$$[-\partial^{2} - 2i(A\partial) + e^{2}A^{2} - m^{2} -
ie(\gamma k)(\gamma A')]\psi = 0. \eqno(11)$$

We are looking for the solution of the last equation in the form:

$$\psi = e^{-ipx}F(\varphi).\eqno(12)$$

After insertion of this relation  into (11), we get with ($k^{2} = 0$)

$$\partial^{\mu}F = k^{\mu}F', \quad \partial_{\mu}\partial^{\mu}F = k^{2}F''
= 0,\eqno(13)$$
the following equation for $F(\varphi)$

$$2i(kp)F' + [-2e(pA) + e^{2}A^{2} - ie(\gamma k)(\gamma A')]F = 0. \eqno(14)$$

The integral of the last equation is of the form:

$$F = \exp\left\{-i\int_{0}^{kx}\left[\frac {e(pA)}{(kp)} - \frac
{e^{2}}{2(kp)}A^{2}\right]
d\varphi + \frac {e(\gamma k)(\gamma A)}{2(kp)}\right\}
\frac{u}{\sqrt{2p_{0}}}, \eqno(15)$$
where $u/\sqrt{2p_{0}}$ is the arbitrary constant bispinor.

Al powers of $(\gamma k)(\gamma A)$ above the first are equal to zero,
since

$$(\gamma k)(\gamma A)(\gamma k)(\gamma A) =
- (\gamma k)(\gamma k)(\gamma A)(\gamma A) +
2(kA)(\gamma k)(\gamma A) = -k^{2}A^{2} = 0.\eqno(16)$$

Then we can write:

$$\exp\left\{e\frac {(\gamma k)(\gamma A)}{2(kp)}\right\} =
1 + \frac {e(\gamma k)(\gamma A)}{2(kp)}.\eqno(17)$$

So, the  solution is of the form:

$$\psi_{p} = R \frac {u}{\sqrt{2p_{0}}}e^{iS}  =
\left[1 + \frac {e}{2kp}(\gamma k)(\gamma A)\right]
\frac {u}{\sqrt{2p_{0}}}e^{iS},
\eqno(18)$$
where $u$ is an electron bispinor of the corresponding Dirac equation

$$(\gamma p - m)u = 0,\eqno(19)$$
with the normalization condition ${\bar u}u = 2m$

The mathematical object $S$ is the classical Hamilton-Jacobi function,
which  was determined in the form:

$$S = -px - \int_{0}^{kx}\frac {e}{(kp)}\left[(pA) - \frac {e}{2}
(A)^{2}\right]d\varphi. \eqno(20)$$

The current density is

$$j^{\mu} = {\bar \psi}_{p}\gamma^{\mu}\psi_{p},
\eqno(21)$$
where $\bar\psi_{p}$ is defined as the transposition of (18), or,

$$\bar\psi_{p} = \frac {\bar u}{\sqrt{2p_{0}}}\left[1 +
\frac {e}{2kp}(\gamma A)(\gamma k)\right]
e^{-iS}.
\eqno(22)$$

After insertion of $\psi_{p}$ and $\bar\psi_{p}$
into the current density, we have:

$$j^{\mu} = \frac {1}{p_{0}}\left\{p^{\mu} - eA^{\mu} +
k^{\mu}\left(\frac {e(pA)}{(kp)} - \frac {e^{2}A^{2}}{2(kp)}\right)
\right\}.
\eqno(23)$$
which is in agreement with formula in the Meyer article [3].

If $A^{\mu}(\varphi)$ are periodic functions, and their time-average value is
zero, then the mean value of the current density is

$$\overline{j^{\mu}} = \frac {1}{p_{0}}\left(p^{\mu} - \frac {e^{2}}{2(kp)}\overline {A^{2}}k^{\mu}
\right) = \frac {q^{\mu}}{p_{0}}.\eqno(24)$$

\section{Volkov solution of the Dirac equation for plane wave
in a dielectric medium}

The mathematical approach to the situation where we consider plane wave
solution in a medium is the same, only with the difference that the Lorentz
condition must be replaced by the following one [4]:

$$\partial_{\mu}A^{\mu} =  kA' = (\mu\varepsilon - 1)(\eta \partial)(\eta A)=
(\mu\varepsilon - 1)(\eta k)(\eta A')
\eqno(25)$$
with the specification $\eta^{\mu} = (1, {\bf 0})$ as the unit time-like vector in the rest
frame of the medium [4].

For periodic potential $A^{\mu}$ we then get from equation (25) instead of $kA = 0$ the
following equation:

$$kA = (\mu\varepsilon - 1)(\eta k)(\eta A). \eqno(26)$$

Then, we get instead of equation (14) the following equation for function
$F(\varphi)$:

$$2i(kp)F' + [-2e(pA) + e^{2}A^{2} - ie(\gamma k)(\gamma A') -
ie(\mu\varepsilon - 1)(\eta k)(\eta A')]F = 0 . \eqno(27)$$

The solution of the last equation is the solution of the linear equation
of the form $y' + Py = 0$ and it means it is of the form $y =
C\exp(-\int Pdx)$, where C is some constant. So,  we can write the
solution as follows:

$$F = \exp\left\{-i\int_{0}^{kx}\left[\frac {e}{(kp)}(pA) - \frac
{e^{2}}{2(kp)}A^{2}\right]d\varphi +
\frac {e(\gamma k)(\gamma A)}{2(kp)} + \frac {e}{2(kp)}\alpha
\right\}
\frac{u}{\sqrt{2p_{0}}}, \eqno(28)$$
where

$$\alpha = (\mu\varepsilon - 1)(\eta k)(\eta A)\eqno(29)$$

The wave function $\psi$ is then the modified wave function (18), which we
can write in te form

$$\psi_{p} = \left[1 + \sum_{n=1}^{\infty}\left(\frac
    {e}{2(kp)}\right)^{n}
(2\alpha)^{n}
(\gamma k)(\gamma A)\right]
\frac {u}{\sqrt{2p_{0}}}e^{iS}e^{T},
\eqno(30)$$
where

$$T = \frac {e}{2(kp)}(\mu\varepsilon - 1)(\eta k)(\eta A), \eqno(31)$$
and where we used in the  last formula the following relation

$$\left[(\gamma k)(\gamma A)\right]^{n} = (2\alpha)^{n}(\gamma k)(\gamma A).
\eqno(32)$$

So, we see that the influence of the medium on the Volkov solution is
involved in $\exp(T)$, where $T$ is given by equation (31) and in the new
term which involves sum of the infinite number of coefficients.

\section{Compton emission of photons by electron}

Now, let us consider electromagnetic monochromatic plane wave which is
polarized in circle. We write the four-potential in the form:

$$A = a_{1}\cos\varphi + a_{2}\sin\varphi,\eqno(33)$$
where the amplitudes $a_{i}$ are the same and mutually perpendicular, or,

$$a_{1}^{2} = a_{2}^{2} = a^{2}, \quad a_{1}a_{2} = 0. \eqno(34)$$

If we use the derived solution (30) of the Dirac equation in a
medium, we get very complicated formalism. To avoid such complications we use
the  original Volkov approach and then we introduce index of refraction
into  definition of a photon in a medium.
Such approach was used in the similar form by Schwinger et al. [4]
in case of the description of the \v Cerenkov radiation with massless
photons and by
author [5] in case of the \v Cerenkov effect with radiative
corrections and
of the \v Cerenkov effect with massive photons [6].
We hope, that this simple method can be used also in case of the
application of the Volkov solution for photons in medium.

Volkov solution for the standard  vacuum situation is of the form:

$$\psi_{p} = \left\{1 + \left(\frac {e}{2(kp)}\right)
[(\gamma k)(\gamma a_{1})\cos\varphi +
(\gamma k)(\gamma a_{2})\sin\varphi]\right\}\frac {u(p)}{\sqrt{2q_{0}}}
 \quad \times$$

$$\exp\left\{-ie\frac{(a_{1}p)}{(kp)}\sin\varphi + ie\frac
{(a_{2}p)}{(kp)}\cos\varphi -iqx\right\}, \eqno(35)$$
where

$$q^{\mu} = p^{\mu} -e^{2}\frac {a^{2}}{2(kp)}k^{\mu}. \eqno(36)$$
because it follows from eq. (24).

We know that the matrix element $M$ corresponding to the emission of
photon by electron in the electromagnetic field is as follows [2]:

$$S_{fi} = -ie^{2}\int d^{4}x \bar \psi_{p'}(\gamma e'^{*})
\psi_{p}\frac {e^{ik'x}}{\sqrt{2\omega'}},
\eqno(37)$$
where $\psi_{p}$ is the wave function of an electron before interaction with
the laser pulse and $\psi_{p'}$ is the wave function of electron after
emission of photon with components $k'^{\mu} = (\omega', {\bf k}')$.
The quantity $e'^{*}$ is the four polarization vector of emitted photon.

Then, we get the following linear combination in the matrix element:

$$e^{-i\alpha_{1}\sin\varphi + i\alpha_{2}\cos\varphi}\eqno(38)$$

$$e^{-i\alpha_{1}\sin\varphi + i\alpha_{2}\cos\varphi}\cos\varphi\eqno(39)$$

$$e^{-i\alpha_{1}\sin\varphi + i\alpha_{2}\cos\varphi}\sin\varphi.\eqno(40)$$
where

$$\alpha_{1} = e\left(\frac {a_{1}p}{kp} - \frac {a_{2}p'}{kp'}\right),
\eqno(41)$$
and

$$\alpha_{2} = e\left(\frac {a_{2}p}{kp} - \frac {a_{2}p'}{kp'}\right).
\eqno(42)$$

Now, we can expand exponential function into the Fourier transformation where
the coefficients of the expansion will be $B_{s}, B_{1s}, B_{2s}$. So we
write:

$$e^{-i\alpha_{1}\sin\varphi + i\alpha_{2}\cos\varphi}
= e^{-i\sqrt{\alpha_{1}^{2} + \alpha_{2}^{2}}\sin(\varphi - \varphi_{0})} =
\sum_{s = -\infty}^{\infty}B_{s}e^{-is\varphi}\eqno(43)$$

$$e^{-i\alpha_{1}\sin\varphi + i\alpha_{2}\cos\varphi}\cos\varphi
= e^{-i\sqrt{\alpha_{1}^{2} + \alpha_{2}^{2}}\sin(\varphi - \varphi_{0})}\cos\varphi =
\sum_{s = -\infty}^{\infty}B_{1s}e^{-is\varphi}\eqno(44)$$

$$e^{-i\alpha_{1}\sin\varphi + i\alpha_{2}\cos\varphi}\sin\varphi
= e^{-i\sqrt{\alpha_{1}^{2} + \alpha_{2}^{2}}\sin(\varphi - \varphi_{0})}\sin\varphi =
\sum_{s = -\infty}^{\infty}B_{2s}e^{-is\varphi}.\eqno(45)$$

The Coefficients $B_{s}, B_{1s}, B_{2s}$ can be expressed by means of 
the Bessel function as follows
[2]:

$$B_{s} = J_{s}(z)e^{is\varphi_{0}}\eqno(46)$$

$$B_{1s} = \frac {1}{2}\left[J_{s+1}(z)e^{i(s+1)\varphi_{0}} +
J_{s-1}(z)e^{i(s-1)\varphi_{0}}\right]\eqno(47)$$

$$B_{2s} = \frac {1}{2i}\left[J_{s+1}(z)e^{i(s+1)\varphi_{0}} -
J_{s-1}(z)e^{i(s-1)\varphi_{0}}\right],\eqno(48)$$
where
the quantity $z$ is defined in [2] through $\alpha$ components as follows:

$$z = \sqrt{\alpha_{1}^{2} + \alpha_{2}^{2}}\eqno(49)$$
and

$$\cos\varphi_{0} = \frac {\alpha_{1}}{z}; \quad \sin\varphi_{0} =
\frac{\alpha_{2}}{z}\eqno(50)$$

Functions $B_{s}, B_{1s}, B_{2s}$ are related one to another as follows:

$$\alpha_{1}B_{1s} + \alpha_{2}B_{2s} = sB_{s},\eqno(51)$$
which follows from the well known relation for Bessel functions:

$$J_{s-1}(z) + J_{s+1}(z) = \frac {2s}{z}J_{s}(z)\eqno(52)$$

The matrix element (37) can be written in the form [2]:

$$S_{fi} = \frac {1}{\sqrt{2\omega'2q_{0}2q_{0}'}}\sum_{s}M_{fi}^{(s)}
(2\pi)^{4}i\delta^{(4)}(sk + q - q'- k'),\eqno(53)$$
where the $\delta$-function involves the law of conservation in the form

$$sk + q = q' + k' , \eqno(54)$$
where, respecting eq. (24)

$$q^{\mu} = p^{\mu} - \frac {e^{2}}{2(kp)}\overline{A^{2}}k^{\mu}. \eqno(55)$$

Using the last equation we can introduce the so called "effective mass"
of electron immersed in the periodic wave potential as follows:

$$q^{2} = m_{*}^{2}; \quad m_{*} = m\sqrt{\left(1 - \frac {e^{2}}{m^{2}}
\overline{A^{2}}\right)}\eqno(56)$$

Formula (56) represents the mass renormalization of an electron mass in the field $A$.
In other words the mass renormalization is defined by the equation

$$m_{\rm phys} = m_{\rm bare} + \delta m \eqno(57)$$
where $\delta m$ follows from eq. (56). The quantity $m_{\rm phys}$ is
the physical mass that an experimenter would measure if the particle
were
subject to Newton's law ${\bf F} = m_{\rm phys}{\bf a}$. In case of
the periodic field of laser, the quantity $\delta m$ has  the finite
value. The renormalization is not introduced here " by hands " but it follows
immediately from the formulation of the problem of electron in the
wave field.
We know the general opinion that renormalization is unavoidable in the
quantum field theory
[7], [8]. On the other hand Schwinger source theory works
without renormalization at all [9].

We can write

$$q^{2} = q'^{2} = m^{2}(1 + \xi^{2}) \equiv m_{*}^{2},\eqno(58)$$
where for plane wave (35) with relations (36)

$$\xi = \frac{e}{m}\sqrt{-a^{2}}.\eqno(59)$$

According to [2] the matrix element in (53) is of the form:

$$M_{fi}^{(s)} = -e\sqrt{4\pi}\bar u(p')\left\{\left(
(\gamma e') - e^{2}a^{2}\frac {(ke')}{2(kp)} \frac {(\gamma k)}{(kp')}
\right)B_{s} + \right.$$

$$e \left(\frac {(\gamma a_{1})(\gamma k)(\gamma e')}{2(kp')} +
\frac {(\gamma e')(\gamma k)(\gamma a_{1})}{2(kp)}\right)B_{1s} + $$

$$\left. e \left(\frac {(\gamma a_{2})(\gamma k)(\gamma e')}{2(kp')} +
\frac {(\gamma e')(\gamma k)(\gamma a_{2})}{2(kp)}\right)B_{2s}\right\}u(p)
\eqno(60)$$

It is possible to show, that the total probability of the emission of photons 
from unit volume in unit time is [2]:

$$W = \frac {e^{2}m^{2}}{4q_{0}}\sum_{s = 1}^{\infty}
\int \frac {du}{(1+u)^{2}}\quad \times $$

$$\left\{-4J_{s}^{2}(z) + \xi^{2}\left(2 + \frac
{u^{2}}{1+u}\right)\left(J_{s+1}^{2}(z) + J_{s-1}^{2}(z) - 2J_{s}^{2}(z)
\right)\right\}, \eqno(61)$$
where

$$u = \frac {(kk')}{(kp')}, \quad u_{s} = \frac {2s(kp)}{m_{*}^{2}}, \quad 
z =2sm^{2}\frac{\xi}{\sqrt{1 + \xi^{2}}}\*
\sqrt{\frac{u}{u_{s}}\left(1 - \frac{u}{u_{s}}\right)}.
\eqno(62)$$

Variables $\alpha_{1,2}$ are to be expressed in terms of variables
$u$ and $u_{s}$  from the equation (62).

When $\xi \ll 1$ (the condition for perturbation theory to be valid), the
integrand (61) can be expanded in powers of $\xi$. For the first term of
the sum $W_{1}$, we get

$$W_{1} \approx \frac {e^{2}m^{2}}{4p_{0}}\xi^{2}
\int_{0}^{u_{1}}\left[2 +  \frac {u^{2}}{1+u} - 4\frac {u}{u_{1}}
\left(1 - \frac {u}{u_{1}}\right)\right]du = $$

$$\frac {e^{2}m^{2}}{4p_{0}}\xi^{2}
\left[\left(1 - \frac {4}{u_{1}} - \frac {8}{u_{1}^{2}}\right)
\ln\left(1 + u_{1}\right) + \frac {1}{2} + \frac {8}{u_{1}} -
\frac {1}{2(1 + u_{1})^{2}}\right]\eqno(63)$$
with

$$u_{1} \approx \frac {2(kp)}{m^{2}}.\eqno(64)$$

It is possible to determine the second and the next harmonics as an analogy
with the Berestetzkii approach, however the aim of this article was only
to illustrate the influence of the dielectric medium on the Compton effect.

Let us consider the equation (54) in the form:

$$sk + q - k' = q'.\eqno(65)$$

The  equation (65) has physical meaning for $s = 1,  2, ...  n$, $n$ being
positive integer. For  $s= 1$ it has meaning of the conservation of energy
momentum of the  one-photon  Compton process,
$s = 2 $ has meaning of the two-photon Compton process and $s = n$
has meaning of the multiphoton interaction with $n$ photons
of laser beam with an electron.  The multiphoton interaction is nonlinear and
differs from the situation where electron scatters twice or more as it
traverses the laser focus. It also means that the
original Einstein photoelectric equation can be replaced by the more general
multiphoton photoelectric equation in the form:

$$n\hbar\omega = \frac {1}{2}mv^{2} + E_{i},
\eqno(66)$$
where $E_{i}$ is the binding energy of the outermost electron in the atomic
system [10].
It means that the ionization
effect occurs also in the case that $\hbar\omega < E_{i}$ in case that
number of participating photons is $n > E_{i}/\hbar \omega$. We will not
solve furthermore this specific problem.

We introduce the scattering angle $\theta$ between ${\bf k}$
and ${\bf k}'$. In other words,
The scattering angle $\theta$ is measured with respect to the
incident photon direction. Then, with  $|{\bf k}| = n\omega$
and  $|{\bf k}'| = n\omega'$, where $n$ is index of refraction
of the dielectric, we get from
the squared equation (65) in the rest system of electron, where
$q = (m_{*},0)$, the following equation:

$$s\frac {1}{\omega'} - \frac {1}{\omega} =
\frac {s}{m_{*}}(1 - n^{2}\cos\theta), \eqno(67)$$
which is modification of the original equation for the Compton process

$$\frac {1}{\omega'} - \frac {1}{\omega} = \frac {1}{m}(1 - \cos\theta).
\eqno(68)$$

So, we see that Compton effect described by the Volkov solution of
the Dirac equation
differs from the original Compton formula only by the existence of the
renormalized mass
and the presentation of index of refraction.

We know that the last formula of the original Compton effect can be written
in the form suitable for the experimental verification, namely:

$$\Delta \lambda = 4\pi\frac{\hbar}{mc}\sin^{2}\frac {\theta}{2},
\eqno(69)$$
which was used by Compton for the verification of the quantum
nature of light [11].

If we consider the Compton process in dielectric, then the last formula goes to
 the following form:

$$\Delta \lambda = 2\pi\frac{\hbar}{mc}(1 - n^{2}\cos \theta).
\eqno(70)$$

It is evident that relation $\lambda' - \lambda \geq 0$ follows from eq. (1).
However, if we put

$$1 - n^{2}\cos\theta \leq 0 \eqno(71), $$
or, equivalently

$$\frac{1}{n^{2}} \leq \cos\theta \leq  1,\eqno(72)$$
then, we see that for some angles determined by eq. (72) the relation
$\lambda' - \lambda \leq 0$ follows.
This surprising  result is the
anomalous Compton effect which is caused by the index of refraction of
the medium. To our knowledge, it was not published in the optical or
particle journals.

The equation $sk + q = q' + k'$ is the symbolic expression of
the nonlinear Compton effect in which several
photons are absorbed at a single point, but only single high-energy photon
is emitted. The second process  where electron scatters twice
or more as it traverses the laser focus is not considered here.
The nonlinear Compton process was experimentally confirmed for instance by
Bulla et al. [12].

The formula (67) can be also expressed in terms of $\lambda$ as follows:

$$s\lambda' - \lambda  = \frac {2\pi s}{m_{*}}(1 - n^{2}\cos\theta)
\eqno(73)$$
where we have put $\hbar = c = 1$.

Formula (73) can be use for the verification of the Compton effect in a
dielectric medium and  on the other hand the index
refraction follows from it in the following form:

$$n^{2} = \frac {1}{\cos \theta}
\left[1 - \frac {m_{*}}{2\pi s}(s\lambda' - \lambda)\right].
\eqno(74)$$

It means, if we know the $\theta, \lambda, \lambda', s, m_{*}$, we are able
to determine index of refraction of some dielectric medium from the Compton
effect. To our knowledge, this method was not published in the optical
journals.

\section{Discussion}

We have considered  the Compton effect in the framework of
the Volkov solution of the Dirac equation assuming
that the process occurred in medium with the index of refraction $n$.
We have determined the Compton formula for such  physical situation,
and we have got the formula from which we can determine the index of
refraction of a dielectric medium. Mass renormalization of electron is
involved in the Volkov solution.

The index of refraction which is substantial in the considered process,
is effective quantity which was introduced into
physics from experiment. Following Lorentz, the contributions of large
numbers of atoms can be averaged to represent the behavior of an
isotropic dielectric medium with the index of refraction $n$.
We know that the introducing the index of refraction as a constant physical
quantity is inadequate, because it fails to account for the decomposition of
white light in prism. In other words index of refraction depends on frequency.
We can understand it only by learning
more about the optical properties of matter. We know that the electric
composition of matter is, that every atom of matter consists of positive
nucleus  and negative electrons. After application of the electric field,
the polarization of the system occurs. Let be ion represented as $\bigcirc$, with mass $M$ and charge $+
e$, and the negative electron represented here as $\bullet$, with
charge $- e$ and mass $m$. The polarized micro-system can be expressed as [13]:

$$(m, -e)\quad\bullet \longleftrightarrow \bigcirc\quad(M, e),\eqno(75)$$
where $\longleftrightarrow$ is the denotation of the separation
of electron from ion. The electron charge is separated from the
ion charge after application of electric field. The phenomenon is called
polarization. In other words polarization means that the electric field
displaces the electrons from their rest position. In the elementary theory of
polarization it is supposed that electrons are bound elastically to
their rest position. So, their motion can be described by the
equation for harmonic oscillator with frequency $\omega_{0}$ in the most
simple case. Using some elementary physics and mathematics of the dispersion
theory, we can derive the known formula for the index of refraction of
matter.

$$n^{2} = 1 + \frac {Ne^{2}}{m\varepsilon_{0}}\;\frac {1}{\omega^{2}_{0} -
\omega^{2}},\eqno(76)$$
where N is the number of dispersion electrons per unit volume,
$\varepsilon _{0}$ is the dielectric constant of matter.

Let us remark that the above definition of the index of refraction is
the classical one. On the other hand,  there is the quantum theory of
the dispersion and of the index of refraction. It is described in many
textbooks of quantum mechanics and quantum optics. The modern aspects can be
found in the Crenshaw article [14].

It is substantial of our approach that electromagnetic wave is transmitted by
the bound electrons and not by the free electrons. In such a way
the process can exist in matter where the electromagnetic wave is diffracted by
existing free electrons. And this is the Compton effect.

In particle physics,
we know the processes at the finite temperature
where the medium is usually
formed by the electron-positron pairs and photon gas at finite temperature
[15]. Formulation and solution of the Compton process at finite
temperature is, of course, possible, however we solve the
problem Compton process in dielectric medium.
We have proved that this problem is physically meaningful, because of the
existence electromagnetic waves in medium and free electrons in medium.
Free electrons in medium  contribute also at the index
of refraction, however this contribution is considered here as very
small.

The interesting result of our article is the derivation  that for some
scattering angles given by eq. (73) there exist the so called anomalous
Compton effect, where the wave lengths of scattered photons are
shorter than the wave lengths of the original photons. To our knowledge
this effect was not published in the physical journals.

The present article is continuation of the author
discussion on laser problems
[16, 17, 18, 19],
where the discussion of the Compton process in the model of the
laser acceleration was considered.

It is obvious that the Compton scattering is, at the present time, the
elementary laboratory problem because for the monochromatic X-rays for
$\lambda  = 1 \AA$ the shift of wave length is several percent. This is quantity which can be easily measured. On
the other hand the Compton wave length shift for the visible light is only
0,01 percent.
It means that the measurement of the Compton
effect for the visible light in the dielectric medium
involves some obstacles and it means that this problem can be solved
only by the brilliant experimental experts.

\vspace{10mm}

{\bf References}

\vspace{10mm}

\noindent
[1] D. M. Volkov,  $\ddot{\rm U}$ber
eine Klasse von L$\ddot{\rm o}$sungen
der Diracschen Gleichung, {\it Zeitschrift f$\ddot{\it u}$r Physik},
{94} (1935) 250.\\[2mm]
[2] V. B. Berestetzkii, E. M. Lifshitz and  L. P. Pitaevskii,
{\it Quantum Electrodynamics}, Moscow, Nauka, 1989., (in Russian). \\[2mm]
[3] J. W. Meyer, Covariant classical motion of electron in a laser beam,\\
{\it Physical Review D: Particles and Fields}, {3}(2) (1971) 621. \\[2mm]
[4] J. Schwinger, W. Y. Tsai and T. Erber, Classical and quantum theory
of synergic synchrotron-\v Cerenkov radiation,
{\it Annals of Physics (New York)}, {96}(2) (1976) 303. \\[2mm]
[5] M. Pardy, The \v Cerenkov effect with radiative corrections,
{\it Physics  Letters B}, {325} (1994) 517.\\[2mm]
[6] M. Pardy, \v Cerenkov effect with massive photons,
{\it International Journal of Theoretical Physics}, {41}(5) (2002)
 887.\\[2mm]
[7] G. 't Hooft, Nobel lecture: Confrontation with infinity,
{\it Reviews of Modern Physics}, {72} No.2  April (2000) 333. \\[2mm]
[8] M. J. G. Veltman, Nobel Lecture: From weak interactions to gravitation,
{\it Reviews of Modern Physics}, {72} No.2  April (2000) 341. \\[2mm]
[9] W. Dittrich, Source methods in quantum field theory,
{\it Fortschritte der Physik}, {26} (1978) 289.\\[2mm]
[10] N. B. Delone and V. P. Krainov, {\it Multiphoton Processes in Atoms},
2nd ed., Springer-Verlag, Berlin, Heidelberg, New York, 2000. \\[2mm]
[11]J. W. Rohlf, {\it Modern Physics from $\alpha$ to $Z^{0}$},
John Wiley \& Sons, Inc. New York 1994. \\[2mm]
[12] C. Bulla {\it et al.},  Observation of nonlinear effects in Compton
scattering, {\it Physical Review Letter}, {76} (1996) 3116. \\[2mm]
[13] A. Sommerfeld, {\it Optics}, Academic Press, New york 10,
N. Y.,  USA, 1954.\\[2mm]
[14] M. E. Crenshaw, Quantum optics of dielectrics in momentum-space,
Optics Communications, {235} (2004) 153.\\[2mm]
[15] M. E. Ternov, V. Ch.  Zhukovskii, V.P. G. Midodashivili and
P. A. Eminov, The anomalous magnetic moment of electron at finite
temperature, Journal of Nuclear Physics, {43} No. 3 (1986) 764\\[2mm]
[16] M. Pardy, The quantum field theory of laser acceleration,
{\it Physics  Letters A}, {243} (1998) 223. \\[2mm]
[17] M. Pardy, The quantum electrodynamics of laser acceleration,
{\it Radiation Physics and Chemistry}, {61} 2001 321.\\[2mm]
[18] M. Pardy, Electron in the ultrashort laser pulse,
{\it International Journal of Theoretical Physics}, {42}(1) 2003 99.\\[2mm]
[19] M. Pardy, Massive photons and the Volkov solution,
{\it International Journal of Theoretical Physics}, {43}(1) 2004 127.\\[2mm]

\end{document}